\documentclass[aps,prd,showpacs,twocolumn,superscriptaddress,nofootinbib,preprintnumbers]{revtex4-1} 

\makeatletter
\def\p@subsection{}
\makeatother

\usepackage{graphicx,amsmath,amssymb,lipsum,bm}

\usepackage[usenames,dvipsnames]{xcolor}
\definecolor{xlinkcolor}{rgb}{0.7752941176470588, 0.22078431372549023, 0.2262745098039215}

\usepackage[colorlinks=true,citecolor=xlinkcolor,linkcolor=xlinkcolor,urlcolor=xlinkcolor, backref=false,pdfborder={0 0 0}]{hyperref}

\usepackage[normalem]{ulem}
\usepackage[utf8]{inputenc} 
\usepackage{mathtools}
\usepackage{aas_macros}
\usepackage{float}

\usepackage{ulem}
\usepackage{diagbox}

\usepackage[dvipsnames]{xcolor}
\usepackage{mathrsfs}

\newcommand{\be}{\begin{equation}}
\newcommand{\ee}{\end{equation}}
\newcommand{\beqa}{\begin{eqnarray}}
\newcommand{\eeqa}{\end{eqnarray}}

\newcommand\fnl{f_{\rm NL}}

\renewcommand\k{{\bm k}}
\newcommand\q{\bm{q}}
\newcommand\x{\bm x}

\newcommand\G{\mathcal{G}_2}

\newcommand{\bseq}{\begin{subequations}}
\newcommand{\eseq}{\end{subequations}}

\usepackage{xspace}
\newcommand{\paperone}{\citetalias{desi1}\xspace}

\usepackage{color}
\definecolor{darkgreen}{RGB}{0,120,0}

\definecolor{darkgreen}{RGB}{0,120,0}

\def\gsim{\raise0.3ex\hbox{$\;>$\kern-0.75em\raise-1.1ex\hbox{$\sim\;$}}}
\def\lsim{\raise0.3ex\hbox{$\;<$\kern-0.75em\raise-1.1ex\hbox{$\sim\;$}}}

\def\beqn#1{\begin{equation}\label{#1}}
\def\eeqn{\end{equation}}

\def\beqa#1{\begin{eqnarray}\label{#1}}
\def\eeqa{\end{eqnarray}}

\def\kmax{{k_\text{max}}}
\def\hMpc{h{\text{Mpc}}^{-1}}
\def\Mpch{h^{-1}{\text{Mpc}}}

\def\Z2{$\mathcal{Z_2}$}

\defcitealias{desi1}{Paper 1}
\defcitealias{desi2}{Paper 2}
\defcitealias{desi3}{Paper 3}

\usepackage{xspace}
\newcommand{\papertwo}{\citetalias{desi2}\xspace}

\newcommand {\ignore}[1]{}

\begin{document}

\preprint{MIT-CTP/5971}


\title{{\Large Reanalyzing DESI DR1:}\\
3. Constraints on Inflation from Galaxy Power Spectra \& Bispectra}

\author{Anton~Chudaykin}
\email{anton.chudaykin@unige.ch}
\affiliation{D\'epartement de Physique Th\'eorique and Center for Astroparticle Physics,\\
Universit\'e de Gen\`eve, 24 quai Ernest  Ansermet, 1211 Gen\`eve 4, Switzerland}
\author{Mikhail M.~Ivanov}
\email{ivanov99@mit.edu}
\affiliation{Center for Theoretical Physics -- a Leinweber Institute, Massachusetts Institute of Technology, 
Cambridge, MA 02139, USA} 
 \affiliation{The NSF AI Institute for Artificial Intelligence and Fundamental Interactions, Cambridge, MA 02139, USA}
\author{Oliver~H.\,E.~Philcox}
\email{ohep2@cantab.ac.uk}
\affiliation{Leinweber Institute for Theoretical Physics at Stanford, 382 Via Pueblo, Stanford, CA 94305, USA}
\affiliation{Kavli Institute for Particle Astrophysics and Cosmology, 382 Via Pueblo, Stanford, CA 94305, USA}

\begin{abstract} 
\noindent
Models of cosmic inflation generically predict a weak but potentially detectable amount of primordial non-Gaussianity (PNG), which can be used to obtain insights into the degrees of freedom during inflation and their interactions. The simplest types of PNG are the local and non-local (equilateral and orthogonal) shapes of the primordial three-point correlators, which are predicted by models with multiple light fields and derivative interactions in single-field inflation, respectively. In this paper we place constraints on local, equilateral, and orthogonal non-Gaussianities using the power spectrum and bispectrum extracted from first public release of the Dark Energy Spectroscopic Instrument (DESI). Our analysis makes use of higher-order clustering information through a consistent effective field theory (EFT) model for both the power spectrum and bispectrum at one-loop order. Using robust scale cuts where the EFT description is valid, we find the following constraints on PNG amplitudes: $f^{\rm loc}_{\rm NL}=-0.1\pm 7.4$, $f^{\rm equil}_{\rm NL}=719\pm 390$, $f^{\rm orth}_{\rm NL}=-200\pm 100$  (at $68\%$ CL). Non-local PNG constraints can be further improved by combining high-redshift DESI with legacy BOSS data and using simulation-based priors on bias parameters, yielding the strongest large-scale structure constraints to date $f^{\rm equil}_{\rm NL}=200\pm 230$, $f^{\rm orth}_{\rm NL}=-24\pm 86$. Our constraint on $f^{\rm loc}_{\rm NL}$ is competitive with the cosmic microwave background (CMB) limit; the combination gives $\fnl^{\rm loc}=-0.0\pm 4.1$, $18\%$ stronger than the CMB only result, which represents the strongest bound on multi-field inflation yet obtained.
\end{abstract}

\maketitle

\section{Introduction}

\noindent Cosmic inflation explains the flatness of our Universe and the origin of primordial fluctuations~\cite{Starobinsky:1980te,Guth:1980zm,Linde:1981mu,Albrecht:1982wi}.
According to this framework, the early Universe underwent a period of accelerated expansion,
during which its geometry was described by a de Sitter spacetime to a very high accuracy.
This expansion smoothed the Universe on large scales, sourcing the observed flatness and seeded the cosmological fluctuations we observe today. The scale-invariant spectrum of these fluctuations inferred from observations~\cite{WMAP:2003elm,Ade:2013zuv,Planck:2018jri} naturally follows from the scaling symmetry of the de Sitter spacetime.

Inflation also provides a natural explanation for the observed adiabatic nature of cosmological fluctuations, \textit{i.e.}\ the fact that the fluctuations of different species share the same initial conditions. 
One approach to explain it is single-field inflation, in which the adiabaticity is a consequence of having only one degree of freedom, the inflaton, which drives inflation and produces the fluctuations. Another option is multi-field inflation, whereupon multiple degrees of freedom were active during inflation. In this case various interaction mechanisms, such as modulated reheating, thermalization, and beyond, can be used to generate the adiabatic initial conditions, e.g.~\cite{Lyth:2001nq,Enqvist:2001zp,Lyth:2002my,Zaldarriaga:2003my}.

The observed Gaussianity of primordial fluctuations~\cite{Planck:2013wtn,Planck:2019kim,Senatore:2009gt,WMAP:2003xez} suggests that primordial fields during inflation were weakly coupled~\cite{Cheung:2007st,Senatore:2016aui}. Inflationary models, however, generically predict a faint, but possibly experimentally detectable amount of primordial non-Gaussianity (PNG) in the initial conditions. In particular, multi-field models 
imply PNG as a consequence of non-linear interaction mechanisms that produce the fluctuations~\cite{Zaldarriaga:2003my,Enqvist:2001zp,Lyth:2001nq}. The characteristic correlation pattern produced by this type of non-Gaussianity is called \textit{local}, and is characterized by an enhanced coupling between small- and long-wavelength modes.

In the case of single-field inflation, one can think about the inflaton fluctuations as Goldstone modes of the spontaneously broken time-translation symmetry, which can be systematically understood in the context of the effective field theory of inflation (EFTI)~\cite{Cheung:2007st,Cheung:2007sv}. In this picture PNG is generically produced via a non-linear realization of the time-translation symmetry. This makes PNG as natural (from the EFT perspective) as other well established phenomena that appear from non-linear realizations of broken symmetries, such as the interactions of pions~\cite{Donoghue_Golowich_Holstein_2023} or classical sound waves~\cite{Esposito:2018sdc}. 

The EFT of single-field inflation predicts two possible correlation patterns, or `shapes' of non-Gaussianity, commonly referred to as equilateral and orthogonal~\cite{Babich:2004gb,Arkani-Hamed:2003juy,Senatore:2004rj,Alishahiha:2004eh,Chen:2006nt,Senatore:2009gt}. These \textit{non-local} shapes are associated with two distinct types of derivative interactions allowed by symmetries and are highly distinct from the local one. This occurs due to the Maldacena consistency condition~\cite{Maldacena:2002vr,Creminelli:2013mca}, which implies that single-field correlations cannot produce 
any observable coupling between small and long-wavelength modes~\cite{Cabass:2016cgp}. Thus, a detection 
of local PNG would automatically rule out single-field inflation.\footnote{Note that the converse is in general not true.
Multi-field models in principle can produce equilateral PNG (including by `integrating-out' heavy modes) \cite{Tolley:2009fg,Green:2009ds,Senatore:2010wk,Langlois:2008mn,Langlois:2008wt,Arroja:2008yy,Cai:2008if,Cai:2009hw,Chen:2009zp,Chen:2009we,Chen:2010xka,Baumann:2011nk,Arkani-Hamed:2015bza,Cabass:2024wob}, so the
detection of this shape will not rule out multi-field inflation.}

The leading observational constraints on PNG come from analyses of the cosmic microwave background (CMB) temperature and polarization anisotropies. In particular, the Planck PR4 limits on the amplitudes $\fnl$
of the above three shapes of non-Gaussianity yield~\cite{Jung:2025nss} (see also \cite{Planck:2019kim}):
\be
\begin{split}
& \fnl^{\rm loc}=-0.1\pm 5.0\,,\\
& \fnl^{\rm equil}=6\pm 46\,,\\
& \fnl^{\rm orth}=8\pm 21.
\end{split} 
\ee 
These limits are still far from the theoretically interesting benchmarks: $\fnl^{\rm loc}\sim 1$ and $\fnl^{\rm equil},\fnl^{\rm orth}\sim 1$. The former allows one to distinguish between single- and multi-field inflationary models, whilst the latter may provide insights into the ultraviolet completion of the EFTI~\cite{Baumann:2011su}.

Future progress in PNG searches will require data from upcoming large-scale structure surveys (LSS). Though still in their infancy, these can provide access to more 
Fourier modes than the CMB experiments~\cite{Sailer:2021yzm,Ferraro:2022cmj,Cabass:2022epm,Bock:2025ijl}, particularly given the coming saturation of the primary CMB (as evidenced by Simons Observatory forecasts \cite{SimonsObservatory:2018koc,SimonsObservatory:2025wwn}). Whilst the single-field PNG constraints from LSS are still quite a bit weaker than the CMB constraints~\cite{Cabass:2022epm,DAmico:2022gki,Chen:2024bdg,Ivanov:2024hgq}, those on local PNG are already approaching the CMB bounds, due to the dominant `scale-dependent bias effect'~\cite{Dalal:2007cu}, as demonstrated in a slew of recent $f_{\rm NL}^{\rm loc}$ measurements~\cite{Cabass:2022ymb,DAmico:2022gki,Rezaie:2023lvi,DESI:2023duv,Cagliari:2023mkq,Chaussidon:2024qni,Bermejo-Climent:2024bcb,Fabbian:2025fdk}.

In this work we present new limits on PNG using the first year clustering data from the Dark Energy Spectroscopic Instrument (DESI). We utilize the independent analysis pipeline developed in~\cite{Chudaykin:2025aux} and~\cite{Chudaykin:2025lww} (hereafter \paperone and \papertwo, respectively) based on the official Data Release 1 (DR1) catalogs \citep{DESI:2025fxa}. The DESI collaboration has already presented limits on $\fnl^{\rm loc}$ from the large-scale power spectra of quasars and luminous red galaxies~\cite{Chaussidon:2024qni}.
In this work we extend the collaboration result in multiple directions. First, we carry out a search for local PNG from DESI's bright galaxy sample and emission line galaxies, which were not used in the previous PNG searches. Second, we extend the analysis to small scales in order to increase the statistical power of our search, utilizing a consistent one-loop theoretical model. Third, we include the DESI bispectrum dataset for the first time (motivated by our previous works~\cite{Cabass:2022wjy,Cabass:2022ymb,Cabass:2024wob}
and other studies \citep[e.g.,][]{MoradinezhadDizgah:2020whw,Hahn:2023udg}). Finally, we carry out the first dedicated search for non-local PNG using DESI data. 

Our paper is organized as follows. We discuss the data and our analysis pipeline in Section~\ref{sec:data}. Section~\ref{sec:theory} reviews the theoretical background on PNG shapes and the modeling of LSS observables, before we present the results of our search in Section~\ref{sec:res}. Finally, we draw conclusions in Section~\ref{sec:concl}. Our main constraints are presented in Tab.\,\ref{tab:params}.

\section{Data}
\label{sec:data}
\noindent The selection of galaxies and quasars used in this work mainly follows that of \paperone,
but with three additional improvements. Firstly, we use lower $k_{\rm min}$ than the previous study, which is especially important for local PNG analyses. Secondly, we extend the bispectrum analysis to smaller
scales, upgrading to a one-loop theoretical model
~\cite{Philcox:2022frc,Bakx:2025pop} (see also \citep{DAmico:2022ukl}). We also add the bispectrum redshift-space quadrupole data following~\cite{Ivanov:2023qzb}. Finally, we also add the high-redshift quasar sample with $0.8 <z<3.1$ (hereafter denoted `QSO-all') following \citep{Chaussidon:2024qni}. In this case, we use only the large scale power spectrum and bispectrum, noting that this dataset principally impacts local PNG analyses. Below, we give a brief overview of our data sets, estimators, and binning. 

\subsection{Samples}
\noindent This work uses three sets of tracers:

\textbf{Galaxies:} Following \citep{DESI:2024aax}, use the non-overlapping BGS, LRG1, LRG2, LRG3, and ELG2 galaxy samples. Here, BGS, LRG, and ELG refer to the Bright Galaxy Survey (BGS) Luminous Red Galaxy (LRG) and Emission Line Galaxy (ELG) subsections. We do not include the low-redshift ELGs ($z<1.1$) due to angular systematics concerns, following \citep{DESI:2024jis}.

\textbf{QSO:}
This includes quasars in the redshift range $0.8<z<2.1$ with $z_{\rm eff}=1.5$ \citep{DESI:2024aax}. Hereafter, we refer to the Galaxies+QSO combination as `base'.

\textbf{QSO-all:}
This is the complete quasar sample that includes high-redshift quasars that were not included in the main DESI full-shape and BAO analyses~\cite{DESI:2024jis}. In total, the sample contains 1,189,129 quasars spanning the redshift range $0.8<z<3.1$, as in \citep{Chaussidon:2024qni}.

We note that the DESI local PNG study \cite{Chaussidon:2024qni} did not use data from the BGS or ELG samples, due to angular systematic concerns. Here, we do not find any strong systematics in ELG (from comparing weighted and unweighted spectra, as well as the low-$k$ limit of the power spectrum), so we add them to our analysis by default. To test this assumption, we also present $f_{\rm NL}^{\rm loc}$ results from only the LRG sample in in Section~\ref{sec:res}.

\subsection{Estimators}
\noindent 
For each sample, we compute power spectra and bispectra using the quasi-optimal estimators included in the \textsc{PolyBin3D} package  \citep{polybin3d,Philcox:2024rqr}. These account for integral constraints, wide-angle effects, and fiber collisions, as discussed in \paperone. In addition to computing window-convolution matrices, we include an approximate mask-deconvolution step, similar to the pseudo-$C_\ell$ approach used in the CMB community \citep{Alonso:2018jzx}. We subtract off the Poisson shot-noise contribution, using the known analytic form for the constant-in-$k$ contributions (unlike in our previous works, which computed this stochastically). Following \paperone, we also compute Gaussian covariance matrices using \textsc{PolyBin3D}, which incorporate both inhomogeneous shot-noise and masking effects.

For the QSO-all sample, we apply redshift-dependent weights to the galaxy fields entering the power spectrum following~\cite{eBOSS:2019sma,Cagliari:2023mkq,Cagliari:2025rqe}. This involves replacing the usual $P_\ell$ estimator with one involving a cross-spectrum between fields weighted by $w_\ell(z)$ and $\tilde{w}(z)$, where
\begin{eqnarray}
    w_0(z) &=& D(z)\left(b_1(z)-\tfrac{1}{3}f(z)\right)\\\nonumber
    w_2(z) &=& \tfrac{2}{3}D(z)f(z)\\\nonumber
    \tilde{w}(z) &=& b_1(z)-p,
\end{eqnarray}
for linear bias $b_1(z)$ (using the fit of \citep{Chaussidon:2024qni}), growth factor $D(z)$ and growth rate $f(z)$. $p$ enters the non-Gaussian bias parameter (see \ref{eq:pbhipr}), and is here set to $1.6$. This weighting enhances the PNG signals of interest, and is additionally propagated to the estimator window matrix (in contrast to \citep{Chaussidon:2024qni}) and Gaussian covariance. Due to the multipole-dependent weights, we obtain different effective redshifts for each statistic, with $z^{P_0}_{\rm eff}=2.05$ and $z^{P_2}_{\rm eff}=1.9$). For the bispectrum, we apply a $z$-dependent weighting to all input fields to ensure that the effective redshift matches that of the power spectrum multipole.

\begin{figure*}[!t]
\includegraphics[width=1\columnwidth]{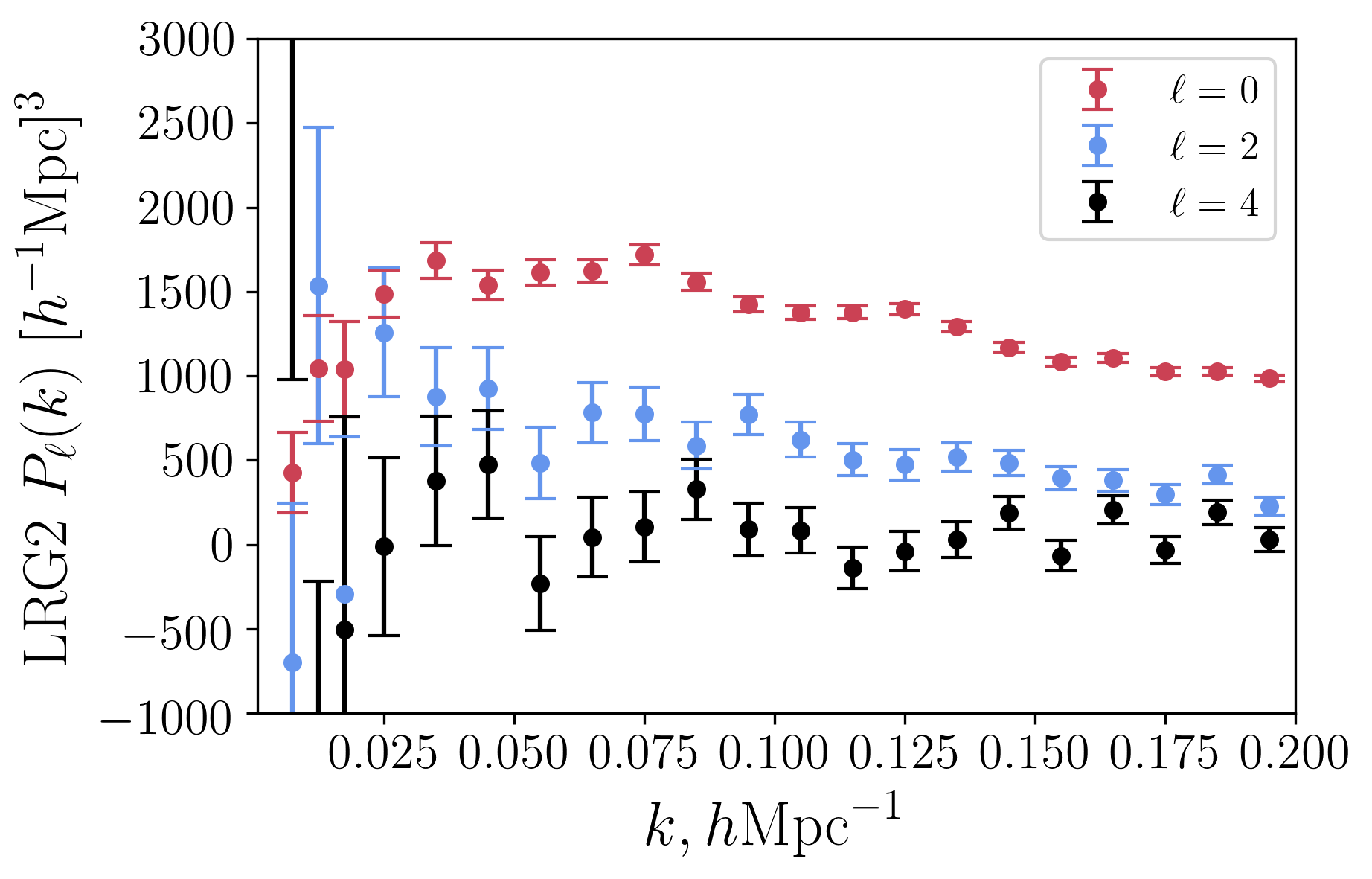}
\includegraphics[width=1\columnwidth]{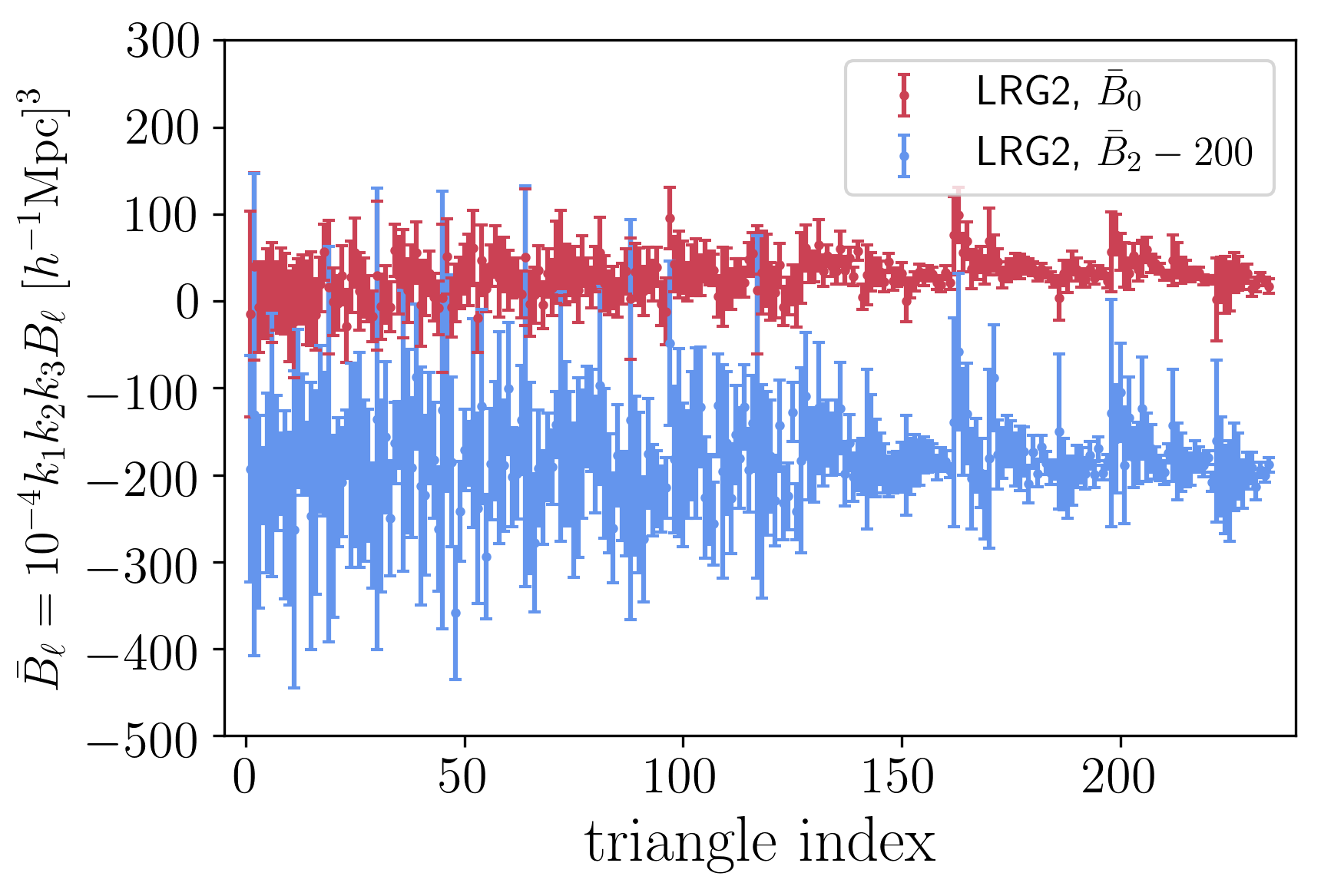}
\includegraphics[width=1\columnwidth]{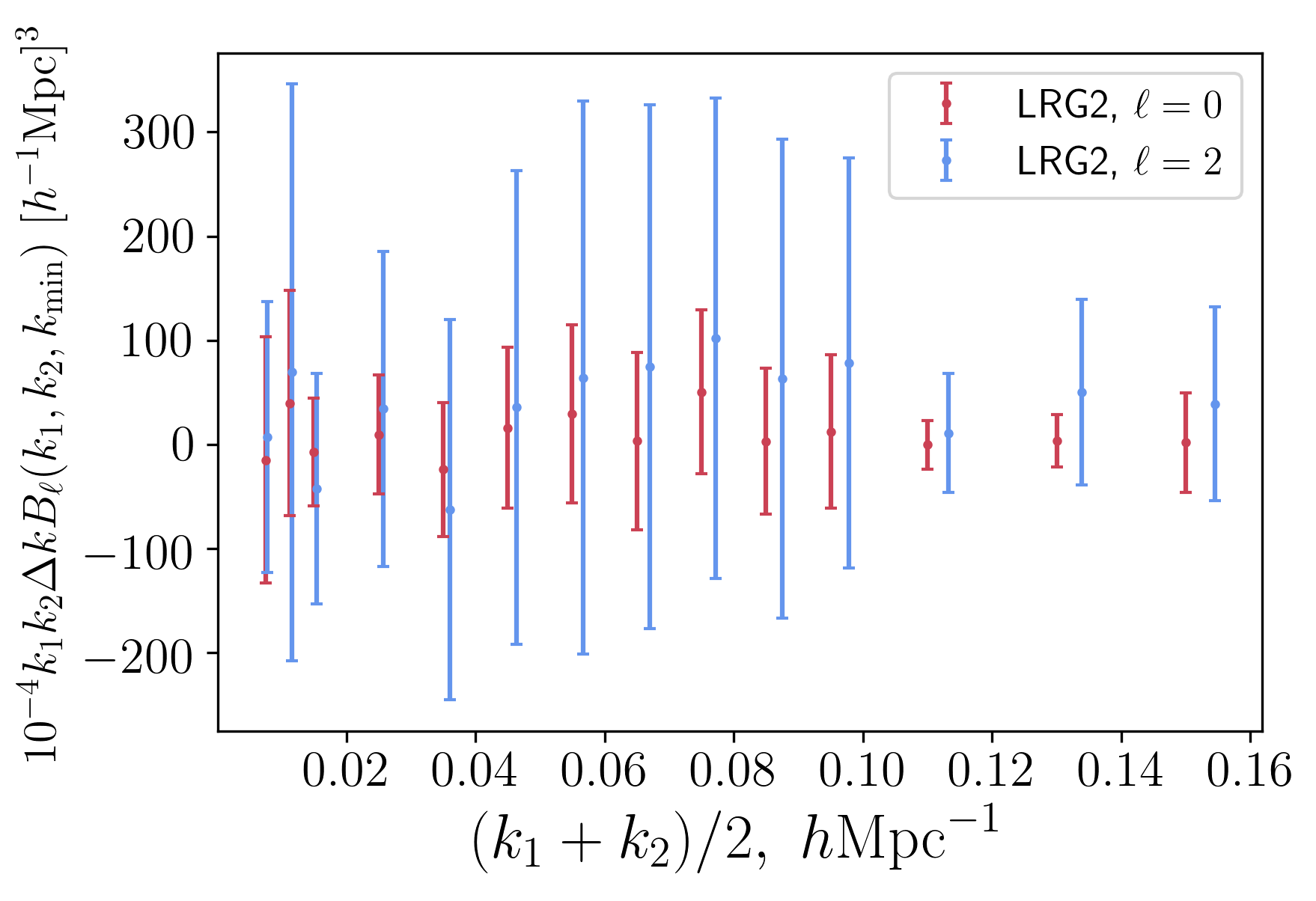}
\includegraphics[width=1\columnwidth]{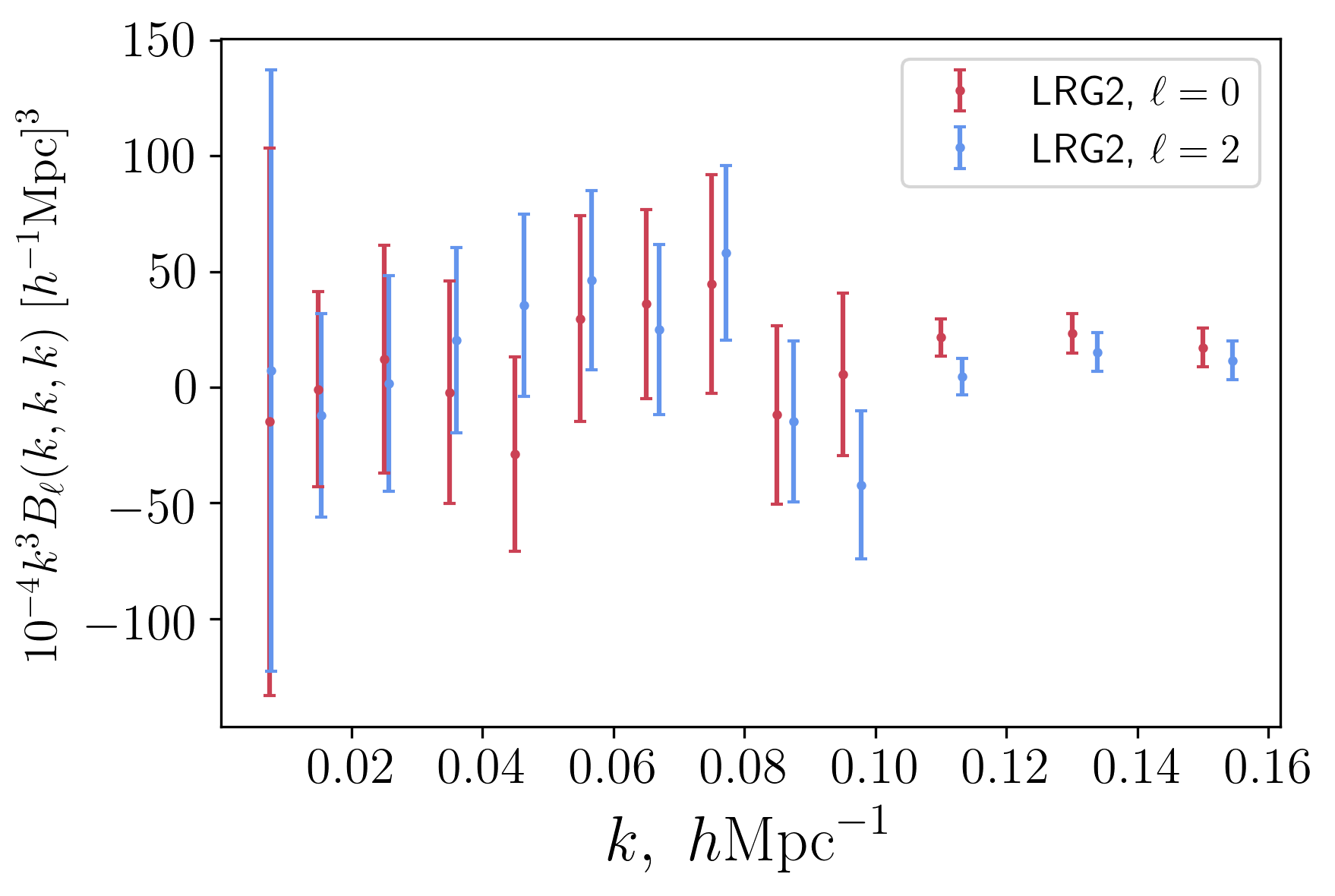}
\caption{\textit{Upper panel:}
Power spectrum and bispectrum measurements for the DESI DR1 LRG2 sample at $z_{\rm eff}=0.7$. The bispectrum quadrupole measurements
are offset vertically for better visibility, and the Poisson shot noise contribution is subtracted from all relevant statistics. The bispectrum dataset includes bins with centers ranging from $k_{\rm min}=0.0075\,\hMpc$ to $\kmax=0.15\,\hMpc$.
\textit{Lower panel:} Bispectrum multipoles for the squeezed ($k_{\rm min} =0.0075\,\hMpc$) and equilateral configurations. The measurements are offset horizontally for better visibility.
}
\label{fig:data_lrg2}
\end{figure*}

\begin{figure*}[!t]
\includegraphics[width=1\columnwidth]{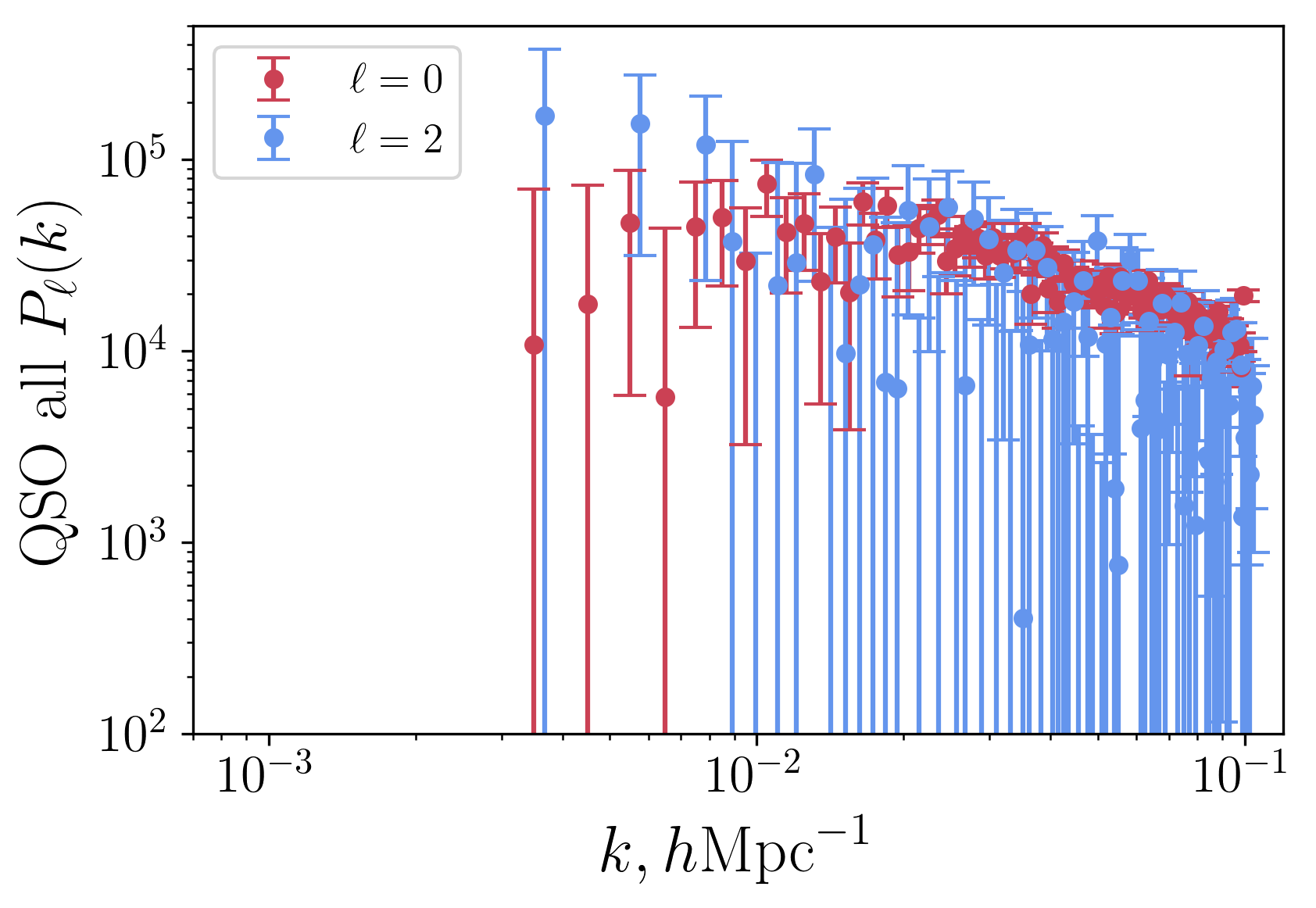}
\includegraphics[width=1\columnwidth]{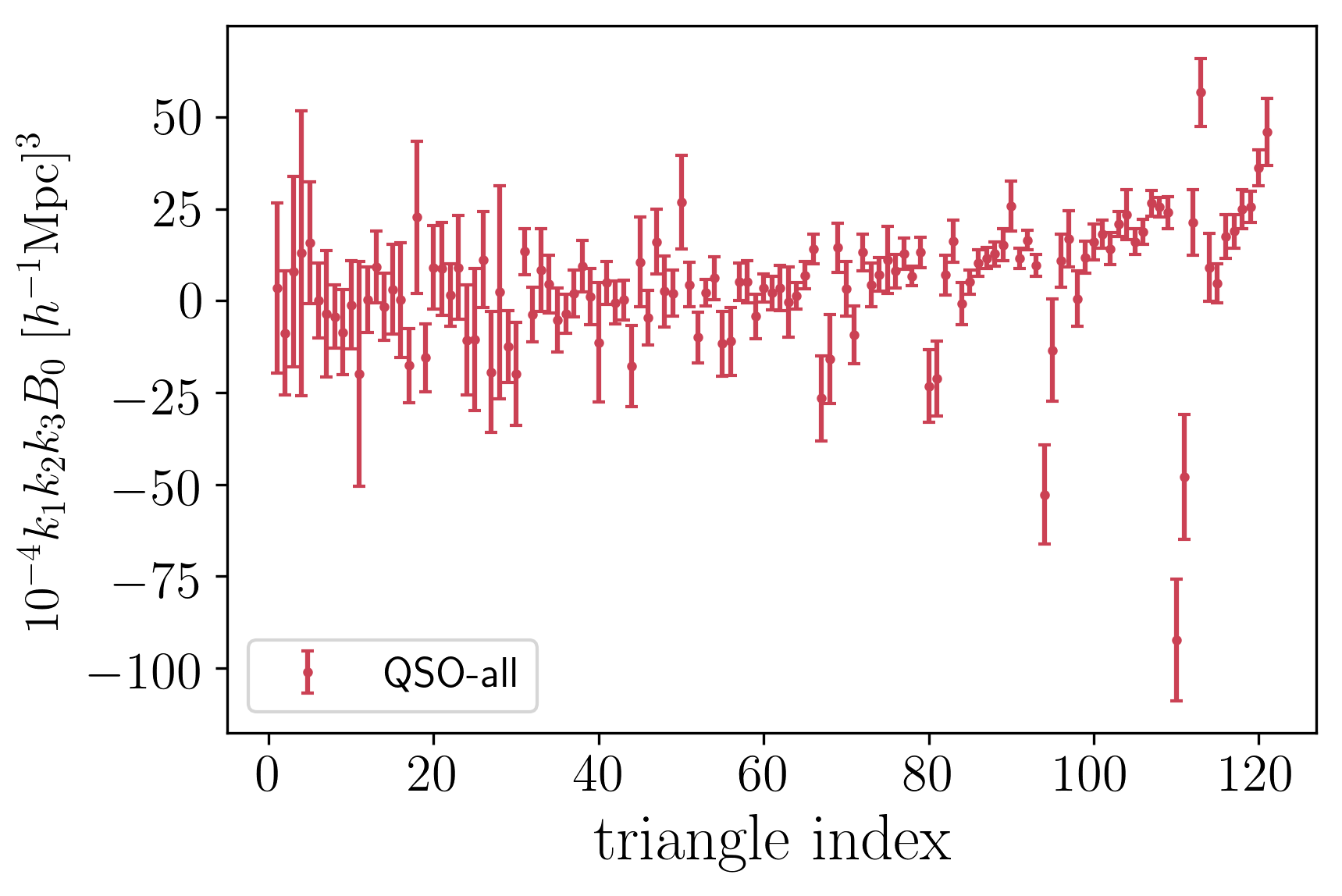}
\includegraphics[width=1\columnwidth]{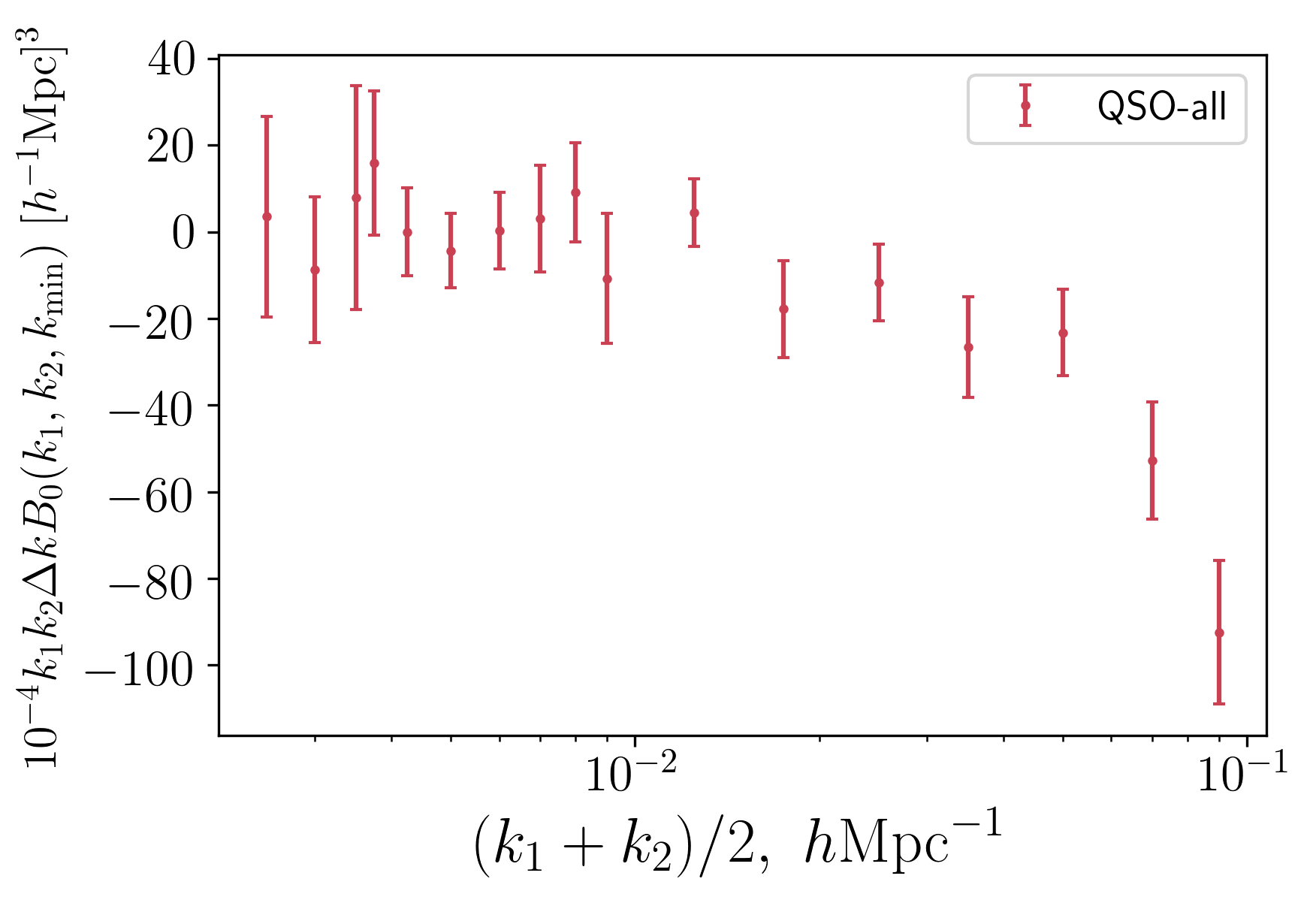}
\includegraphics[width=1\columnwidth]{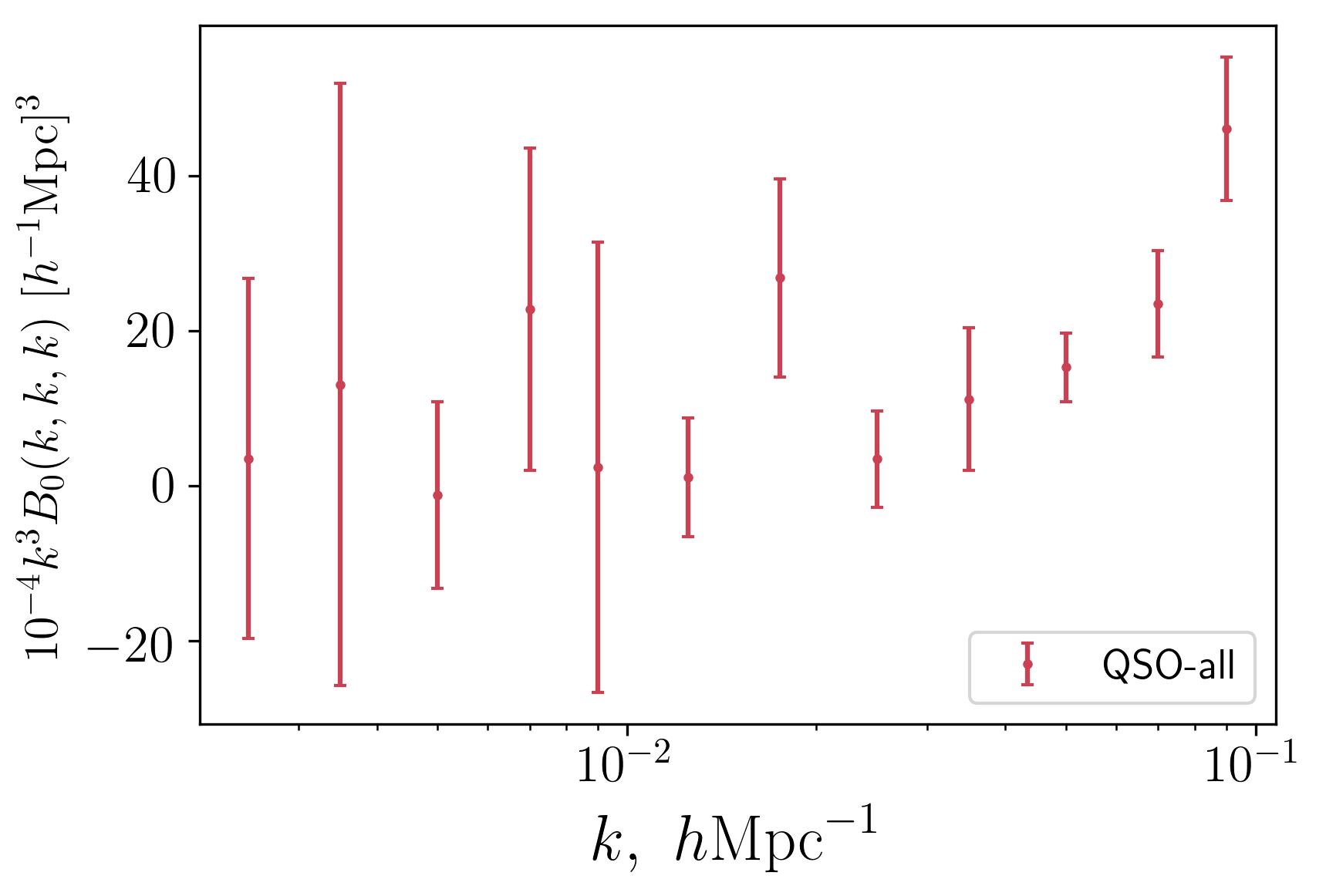}
\caption{As Fig.~\ref{fig:data_lrg2}, but for the QSO-all sample (including quasars up to $z=3.1$). Here, we set $k_{\rm min} =0.0025\,\hMpc$ for the squeezed bispectra in the lower left panel. Since the primary use-case of this sample is for local PNG studies, we omit the power spectrum hexadecapole and bispectrum quadrupole for this sample, and restrict to $k\leq 0.10\,\hMpc$.}
\label{fig:data_qso_all}
\end{figure*}

\subsection{Binning}
\noindent For both galaxies and QSO, we include the power spectrum multipoles, $P_\ell(k)$, with $\ell=0,2,4$, as well as the bispectrum multipoles, $B_\ell(k_1,k_2,k_3)$ for $\ell=0,2$ 
following~\cite{Ivanov:2023qzb}. We use the following scale cuts in our baseline analysis: $k^{P_\ell}_{\rm min}=k^{B_\ell}_{\rm min}=0.005\,\hMpc$, $k^{P_\ell}_{\rm max}=0.2\,\hMpc$, $k^{B_\ell}_{\rm max}=0.16\,\hMpc$. 
In the 
tree-level bispectrum
analyses 
we use $k^{B_\ell}_{\rm max}=0.08\,\hMpc$.
These have been calibrated against various suites of simulations~\cite{Ivanov:2019pdj,Chudaykin:2020ghx,Chudaykin:2020hbf,Ivanov:2021kcd,Ivanov:2021zmi,Chudaykin:2022nru,Ivanov:2023qzb,Chudaykin:2024wlw,Bakx:2025pop} and ensure negligible 
systematic error due to higher-order corrections. 
This is particularly important for the non-Gaussianity analysis, with \cite{Bakx:2025pop} finding that the one-loop EFT 
bispectrum model introduces a bias as large as $\Delta \fnl^{\rm equil}\approx 500$ for $k_{\rm max}^{B_\ell} \geq 0.17\,\hMpc$. The scale cuts used in this work are chosen to minimize such biases.

For QSO-all, we restrict to the large-scale power spectrum monopole and quadrupole, and the bispectrum monopole, with  the scale cuts $k^{P_\ell}_{\rm max}=k^{B_\ell}_{\rm max}=0.1\,\hMpc$, 
and $k^{P_\ell}_{\rm min}=k^{B_\ell}_{\rm min}=0.002\,\hMpc$. This is chosen for consistency with \citep{Chaussidon:2024qni} and will ensure tight $f_{\rm NL}^{\rm loc}$ constraints, noting that there is negligible signal-to-noise in the higher-order bispectrum multipoles on these scales. 

To maximize sensitivity to local PNG while keeping the size of the data-vector manageable, we use an inhomogeneous binning scheme. For the power spectrum, we use bin width $\Delta k = 0.005\,\hMpc$ up to $k=0.02\,\hMpc$ and $\Delta k = 0.01\,\hMpc$ thereafter. For the bispectrum, we use $\Delta k = 0.005\,\hMpc$ up to $k=0.01\,\hMpc$, $\Delta k = 0.01\,\hMpc$ up to $k=0.1\,\hMpc$ and $\Delta k = 0.02\,\hMpc$ onward. This choice allows us to both retain good resolution on large scales and have a relatively small number
of data bins on small scales. With our baseline scale cuts this results in 234 triangle configurations per sample.

For the QSO-all sample, we use 98 $k$ bins for the power spectrum linearly spaced with $\Delta k=0.001\,\hMpc$. For the bispectrum monopole, we use all the closed triangles (triplets of $k$ bins whose centers satisfy momentum conservation), spanning inhomogeneous-size 
bins with the following 
edges 
:
\be 
\begin{split}
    [&0.002  , 0.003  , 0.00425, 
    0.006,  0.008, 
     0.01075, 0.015, \\ & 
       0.02125, 0.03   , 0.0425 ,  0.06   , 0.08   , 0.1    ]\,\hMpc.
\end{split}
       \ee

Our measurements of the LRG2 power spectrum and bispectrum multipoles are shown in Fig.~\ref{fig:data_lrg2}. The results for the other DESI samples are similar (and are included in \paperone). The corresponding measurements for the QSO-all sample are shown in Fig.~\ref{fig:data_qso_all}.

\section{Theoretical Modeling}
\label{sec:theory}

\noindent
In this section we give theoretical background on the shapes of PNG analyzed herein, and then discuss the modeling of the 
anisotropic redshift-space galaxy clustering, including the modifications induced by PNG. 

\subsection{PNG shapes}
\noindent The simplest PNG models produce a non-trivial three-point 
correlation of the cosmological primordial gravitational (Bardeen)
potential $\phi$,
\be 
\langle \phi(\mathbf{k}_1)\,\phi(\mathbf{k}_2)\,\phi(\mathbf{k}_3) \rangle
\equiv B_\phi(k_1,k_2,k_3) (2\pi)^3\,\delta_D^{(3)}(\mathbf{k}_{123}).
\label{eq:bispectrum_def}
\ee
Neglecting the small tilt of the primordial power spectrum, 
the bispectrum can be parameterized in terms of its amplitude 
$\fnl$ and shape $\mathcal{S}$ as
\be 
    B_\phi(k_1, k_2, k_3) = 6 A_\phi^2 f_{\rm NL} \frac{\mathcal{S}(k_1, k_2, k_3)}{k_1^2 k_2^2 k_3^2} .
\ee
where $P_\phi(k) =  A_\phi/k^3$ is the power spectrum. The local shape is given by
\be 
S_{\mathrm{loc}}(k_1, k_2, k_3) = 
\frac{1}{3}\,\frac{k_1^2}{k_2 k_3} + 2~\mathrm{perms.} 
\ee
while the non-local single-field PNG shapes read
\cite{Senatore:2009gt, Babich:2004gb}
\begin{align}
S_{\mathrm{equil}}(k_1, k_2, k_3) 
&= \left( \frac{k_1}{k_2} + 5~\mathrm{perms.} \right) \notag\\
&\quad - \left( \frac{k_1^2}{k_2 k_3} + 2~\mathrm{perms.} \right) - 2 , \label{eq:Sequil}\\[6pt]
S_{\mathrm{ortho}}(k_1, k_2, k_3) 
&= (1 + \bar{p})\,\frac{\Delta}{e_3}
- \bar{p}\,\frac{\Gamma^3}{e_3^2} , \label{eq:Sortho}
\end{align}
where $\bar{p} \approx 8.52587$,
\begin{align}
\Delta &= (k_{123} - 2k_1)(k_{123} - 2k_2)(k_{123} - 2k_3), \nonumber \\[4pt]
k_{123} &= k_1 + k_2 + k_3, \quad 
e_2 = k_1 k_2 + k_2 k_3 + k_1 k_3, \nonumber \\[4pt]
e_3 &= k_1 k_2 k_3, \quad 
\Gamma = \frac{2}{3} e_2 - \frac{1}{3} (k_1^2 + k_2^2 + k_3^2) .
\end{align}
Note that the local shape peaks in the squeezed limit ($k_1/k_2\to 0$),
\be \label{eq: squeezed-local}
\lim_{k_1/k_2\to 0} B_{\phi}^{\mathrm{loc}}(k_1,k_2,|\k_1+\k_2|) = 4 f_{\rm NL}^{\mathrm{loc}} P_\phi(k_1)\,P_\phi(k_2)\,,
\ee 
which can be contrasted with the behavior of single-field PNG:
\be \label{eq: squeezed-single}
\begin{split}
& \lim_{k_1/k_2\to 0} B_{\phi}^{\mathrm{equil,orth}}(k_1,k_2,|\k_1+\k_2|) \\
&\propto \left(\frac{k_1}{k_2}\right)^2 f_{\rm NL}^{\mathrm{equil,orth}} P_\phi(k_1)\,P_\phi(k_2).
\end{split}
\ee 
Note that the orthogonal template used here has the correctly 
suppressed squeezed limit that matches the physical prediction based on EFTI~\cite{Senatore:2009gt}. This can be contrasted with the  approximate  ``orthogonal'' template used in CMB analyses~\cite{Planck:2019kim}, which scales as $(k_1/k_2)$ in the squeezed limit. The difference in the scaling limit between local and non-local PNG will lead to profound differences at the level of structure formation, as we discuss below.

\subsection{LSS with Gaussian Initial Conditions}
\noindent 
The search for non-Gaussian correlations in LSS data is complicated by intrinsic non-linearity induced by dark matter collapse and galaxy formation physics. These processes couple to the Gaussian part of the cosmological fluctuations and produce non-Gaussian contributions to the observed galaxy power spectrum and bispectrum. They can be thought of as a Gaussian background to the LSS observables, which must be modeled to very high precision to search PNG signatures.

The basis for our EFT description of galaxies in redshift space is the perturbative expansion of the galaxy over-density field $\delta_{\rm g}$ in terms of the linear matter density $\delta^{(1)}$ in cosmological perturbation theory. At the leading order this reduces to the standard Eulerian perturbation theory expansion~\cite{Bernardeau:2001qr,Baumann:2010tm,Carrasco:2012cv,Perko:2016puo,Ivanov:2022mrd}, 
\be 
\begin{split}
\delta_{\rm g}(\k) &= \sum_{n=1}\delta_n(\k)\\\nonumber
&= 
\sum_{n=1}\left[\prod^n_{m=1}\int_{\q_m}
\delta^{(1)}(\q_m) \right]
\\
&\quad\times\,(2\pi)^3\delta_D^{(3)}(\k-\sum_{i=1}^n\q_i)
Z_n(\q_1,...,\q_n) ~\,,
\end{split}
\ee 
where $Z_n$ are non-linear kernels~\cite{Assassi:2014fva,Assassi:2015fma,Assassi:2015jqa,Desjacques:2016bnm}. In EFT the above expansion is corrected by the counterterms that capture the backreaction of small scales. 

In the absence of PNG $\delta^{(1)}$ has only Gaussian correlations, \textit{i.e.}\ 
\be 
\langle\delta^{(1)}(\k) \delta^{(1)}(\k')\rangle' = P_{\rm lin}(k)
\ee 
where primes denote correlators stripped of Dirac delta functions and factors of $(2\pi)^3$. The Gaussian one-loop power spectrum
model reads~\cite{Baumann:2010tm,Carrasco:2012cv,Assassi:2014fva,Baldauf:2014qfa,Baldauf:2015aha,Desjacques:2016bnm,Perko:2016puo,Ivanov:2019pdj,Chudaykin:2020aoj}
\be 
P_{\rm gg}= P_{11}(k) + P_{22}+P_{13}+P_{\rm ctr}
+P_{\rm stoch}~\,,
\ee 
where $P_{mn}=s_{mn}\langle 
\delta_m(\k)
\delta_n(\k')\rangle'$, and $s_{mn}$ is a combinatorial factor, e.g., $s_{13}=2$. Here, $P_{22}$ and $P_{13}$ are mode-coupling one-loop contributions and $P_{\rm ctr}$ is the counterterm contribution, which captures the higher-derivative bias, small-scale backreaction from collapsed structures, 
baryonic feedback, and fingers-of-God in redshift space. 
Finally, $P_{\rm stoch}$ is the stochastic contribution
that describes the part of the galaxy density field
uncorrelated with the cosmological initial conditions. 

We can similarly derive the contribution to the one-loop bispectrum from Gaussian initial conditions~\cite{Philcox:2022frc,DAmico:2022ukl,Bakx:2025pop} (building on a number of previous works~\cite{Scoccimarro:1997st,Scoccimarro:1999ed,Scoccimarro:2000sn,Scoccimarro:2000sp,Sefusatti:2006pa,Sefusatti:2007ih,Sefusatti:2009qh,Baldauf:2014qfa,Eggemeier:2018qae,Eggemeier:2021cam,Ivanov:2021kcd,Philcox:2021kcw}):
\be 
\begin{split}
&B_{\rm ggg}=B_{211}+B_{411}+B_{222}+B_{321}^{I} + B_{222}+B_{321}^{II}~\\
& + B_{\rm stoch}+B_{\rm ctr}+B_{\rm mixed}^{\rm tree} + B_{\rm mixed}^{\rm 1-loop} + B_{\rm mixed}^{\rm ctr} ~\,,
\end{split}
\ee 
where $B_{\rm mixed}$ are new terms that appear due to the mixing between the stochastic and deterministic density field components,
and $B_{abc}=s_{abc}\langle 
\delta_a(\k_1)
\delta_b(\k_2)
\delta_c(\k_3)\rangle'$.

This theory includes 45 parameters (\textit{i.e.}\ Wilson coefficients) that capture galaxy bias, non-linear redshift space distortions, and galaxy stochasticity.
Our priors on the Gaussian EFT parameters follow~\cite{Bakx:2025pop} for the bispectrum 
and \paperone for the power spectrum,
with the exception of $b_2$ and $b_{\mathcal{G}_2}$, for which we use priors that respect the perturbativity of EFT,\footnote{In \papertwo, we introduced new priors to mitigate prior volume effects in analyses of alternative cosmological models (based on \citep{Tsedrik:2025hmj}). These are not necessary here, since we work at fixed cosmology.} 
\be 
b_2\sigma^2_8(z) \sim \mathcal{N}(0,1)\,, \quad 
b_{\mathcal{G}_2}\sigma^2_8(z) \sim \mathcal{N}(0,1).
\ee 

We also perform an analysis that implements simulation-based
priors for LRG and ELG samples. Specifically, we use the decorated HOD model for LRG~\cite{Zheng:2004id,Zheng:2007zg,Yuan:2022rsc} 
and the decorated high mass quenched model for ELG~\cite{DESI:2023ujh} to generate the priors. 
Our prior generation approach closely follows~\cite{Ivanov:2024xgb,Ivanov:2024dgv}.
In particular, we utilize simulation-based priors only
for the one-loop EFT power spectrum parameters and the cubic bias coefficient $b_3$ as in~\cite{Ivanov:2024xgb,Ivanov:2024dgv}. In the nomenclature of~\cite{Bakx:2025pop}
these are:~\footnote{Note that 
~\cite{Ivanov:2024xgb,Ivanov:2024dgv}
use a slightly parameterization
$b_4\equiv \tilde{c}$.
We also absorb factors of 
$k_{\rm NL}$ in eq.~\eqref{eq:cs2}
into the counterterms to match the notation of~\paperone. 
} 
\be 
\begin{split}
\{
b_1,b_2,b_{\G},b_3,b_{\Gamma_3},
b_{\nabla^2\delta},e_1,c_1,c_2,\tilde{c},P_{\rm shot},a_0,a_2
\}.
\end{split}
\ee 
Note that the field-level technique used in~\cite{Ivanov:2024xgb,Ivanov:2024dgv}
determined only 
the following 
combination of $e_1$
and $c_1$:
\be 
\label{eq:cs2}
c_{s2}=e_1-\frac{1}{2}c_1 f\,,
\ee 
which appears in the 
one-loop power spectrum
counterterm contribution,
\be 
\begin{split}
\delta_{\rm ctr}=
k^2 & \Big(-b_{\nabla^2\delta} +\left(e_1 -\frac{1}{2}c_1 f\right)f\mu^2 
\\
&
-\frac{1}{2}c_2 f^2\mu^4\Big) \delta^{(1)}(\k)\,,
\end{split}
\ee 
where $\mu=(\hat{\bm z}\cdot {\bm k})/k$, and $\hat{\bm z}$
is the line-of-sight unit vector.
To propagate the distribution
of $c_{s2}$ correctly, we impose a correlated
prior on a set $(e_1,c_1)$ in which 
for every $c_{s2}$ realization we draw a sample 
of $c_1$ from a 
conservative Gaussian prior
consistent with~\paperone:
\be 
\frac{c_1}{[\Mpch]^2}\sim \mathcal{N}(0,30^2).
\ee 
$e_1$ is then determined
from Eq.~\eqref{eq:cs2} 
for each sample, 
ensuring that $c_{s2}$ matches the field level
measurements. For the remaining one-loop bispectrum EFT parameters, we always use the conservative priors similar to the ones from \cite{Bakx:2025pop}.

For the LRG samples, we implement the Gaussian approximation to the prior density following~\cite{Chen:2025jnr} for the analytically 
marginalized parameters, and the normalizing-flow model for the explicitly sampled parameters $b_1$, $b_2$ and $b_{\mathcal{G}_2}$
from~\cite{Ivanov:2024hgq}. We leave the implementation of simulation-based priors for quasars along the lines of~\cite{Ivanov:2025qie} for future work. 

\subsection{LSS with PNG}
\noindent 
Primordial non-Gaussianity affects structure formation through the initial conditions. First, it produces a non-vanishing
three-point correlator of $\delta_1$, which generates new non-linear corrections 
to the galaxy power spectrum and bispectrum. In this work we include such corrections at the first non-vanishing order in $f_{\rm NL}$, \textit{i.e.}\ $P_{12}$ and  $B_{111}$~\cite[e.g.,][]{Cabass:2022epm,Cabass:2022ymb,Sefusatti:2009qh,MoradinezhadDizgah:2020whw,DAmico:2022gki}. For the bispectrum, this is a tree-level contribution, while for the power spectrum, it is a a one-loop term.\footnote{In principle, the PNG loop corrections $B^{(I)/(II)}_{221}$, $B^{(I)/(II)}_{311}$~\cite{Assassi:2015jqa}
should also be included in the analysis. Since they appear only at next-to-leading order, their effect is suppressed with respect to both the leading order non-Gaussian terms and the one-loop Gaussian corrections. As such, it is justified to ignore them in our work.} 

PNG also sources new terms in the galaxy bias expansion. Physically, this is caused by the initial correlations between long and short-wavelength modes that modulate galaxy formation. This is controlled by the squeezed limit of the bispectrum, which differs for the local and single-field PNGs. Physically, we can capture the effect using a new field $\psi$ in the bias expansion~\cite{Schmidt:2010gw,Assassi:2015jqa,Assassi:2015fma,Desjacques:2016bnm,MoradinezhadDizgah:2019xun,MoradinezhadDizgah:2020whw}\be 
\psi(\k) \equiv ({k/k_{\rm NL}})^\Delta \phi(\k)\,,
\ee 
where $\phi$ is the primordial Bardeen potential, $\Delta$ is the power-law index of the soft momentum in the squeezed bispectrum,
and $k_{\rm NL}$ is  the normalization scale which is 
typically taken to  match the non-linear
scale of cosmological perturbation theory.
In what follows we set $k_{\rm NL}=0.45\,\hMpc$ as in~\cite{Cabass:2022epm,Sharma:2025xss}.

For the equilateral and orthogonal PNG $\Delta=2$ (as in \ref{eq: squeezed-single}), such that the above contribution scales like a higher derivative bias term on quasi-linear scales. Given this momentum suppression, it is sufficient to include $\psi$ in the bias expansion only at first non-vanishing order:
\be 
\delta_g\Big|_{\rm non-local~ PNG} = b_\psi \fnl \left(\frac{k}{k_{\rm NL}}\right)^2\phi\,,
\ee 
where $b_\psi$ is a free bias parameter. In this 
work we marginalize over $b_\psi$ over a Gaussian prior $b_\psi\sim \mathcal{N}(0,5^2)$ 
following~\cite{Schmidt:2010gw,Baumann:2012bc,Green:2023uyz,Cabass:2022epm,Sharma:2025xss}.

For local PNG $\Delta=0$ (as in \ref{eq: squeezed-local}), and $\psi=\phi$ becomes important already on large scales. Due to the enhancement by $1/k^2$ compared to the single-field PNG, higher order contributions in the PNG bias expansion become important on large scales as well. Specifically, at second order 
one has to include two terms~\cite{Desjacques:2016bnm}:
\be 
\label{eq:lpng}
\delta_g\Big|_{\rm local~PNG} = b_\phi \fnl^{\rm loc} \phi(\q) + b_{\delta \phi}
\fnl^{\rm loc} \delta(\x) \phi(\q)\,, 
\ee 
where $\q$ and $\x$ are Lagrangian and Eulerian coordinates, 
respectively. The above expansion generates additional terms
in the PNG tree-level prediction, 
\be 
P_{\rm gg}^{\rm tree}\Big|_{\rm local~PNG}=\left(2Z_1(\k)+\frac{b_\phi \fnl^{\rm loc} }{\mathcal{M}(k)}\right) \frac{b_\phi 
\fnl^{\rm loc} 
P_{\rm lin}(k)}{\mathcal{M}(k)}\,,
\ee 
where $Z_1(\k)=(b_1+f\mu^2)$,
as well as the loop 
corrections
\be 
P_{\rm gg}^{\rm 1-loop}\Big|_{\rm local~PNG}=P_{12}+P^{\fnl}_{22}+P^{\fnl}_{13}~\,,
\ee 
\cite{Cabass:2022ymb,Assassi:2015fma,MoradinezhadDizgah:2020whw}, where in the last line we have explicitly added the $P_{12}$ contribution for completeness. In addition, \eqref{eq:lpng} generates a new tree-level bispectrum contribution $B_{112}^{\fnl}$~\cite{Cabass:2022ymb}. 

Unlike the non-local PNG, the scale-dependent bias~\eqref{eq:lpng} is the main source of local PNG constraints.
In this case, it has become standard practice to place strong priors on $b_\phi$
and $b_{\delta \phi}$ from semi-analytic arguments or simulations, noting the strong degeneracy between $b_\phi$ and $f_{\rm NL}$.
Whilst the semi-analytic `universality' relations work quite 
well for dark matter halos~\cite{LoVerde:2007ri,2013PhRvD..88b3515S,Scoccimarro:2011pz,Biagetti:2016ywx,Desjacques:2016bnm, Goldstein:2024bky,Dalal:2007cu,djs_11_lpng_bias,desjacques_fnl_review,Assassi:2015fma,baldauf_gr,quijote_png_halo,hadzhiyska_abacuspng,Goldstein:2024bky,Sullivan:2024jxe},
they perform rather poorly for galaxies 
in hydrodynamical simulations~\cite{Barreira:2020ekm,Barreira:2021ukk},
which motivates the use of simulation-based
relations.\footnote{In the strict EFT-sense, one would marginalize over $b_\phi,b_{\phi\delta}$ with broad priors. Due to the almost-perfect degeneracy with $f_{\rm NL}$ (up to loops and bispectra), this severely reduces the information content on local PNG, which motivates fixing these parameters.} Specifically, for the linear bias we use 
\be \label{eq:pbhipr}
b_\phi = 2\delta_c (b_1-p)\,,
\ee 
where we choose $p=0.55$ for BGS and LRG samples, as appropriate for magnitude limited samples~\cite{Barreira:2020ekm}, $p=1$ for ELG2 following the universality relation~\cite{Desjacques:2016bnm}, 
and $p=1.6$ for QSO and QSO-all following from the recent merger models~\cite{Slosar:2008hx,2024ApJ...963...91B} (matching \citep{Chaussidon:2024qni}). For $b_{\delta \phi}$ we use the dark matter halo bias fit from simulations~\cite{Barreira:2021ueb}
\be \label{eq:bdphi}
b_{\delta \phi} = 3.85 -9.49 b_1 +3.44 b_1^2 
\ee 
noting that its predictions are quite close to the universality prediction~\cite{Desjacques:2016bnm,Cabass:2022ymb}.

Finally, we note that for the QSO-all sample, we use the tree EFT computations both for the power spectrum and bispectrum~\cite{Ivanov:2021kcd,Philcox:2021kcw,Cabass:2022wjy}. These are appropriate for the large scales ($\kmax=0.1\,\hMpc$) used in this analysis. 

\section{Results}
\label{sec:res}

\begin{table*}[!t]
    \centering
    \begin{tabular}{lclclcl}
        \hline
        Parameter & \multicolumn{2}{c}{$\fnl^{\rm loc}$} & \multicolumn{2}{c}{$\fnl^{\rm equil}$} & \multicolumn{2}{c}{$\fnl^{\rm orth}$} \\
        \hline
        Galaxies+QSO-all  & $-0.1_{-7.4}^{+7.4}$ && $719_{-400}^{+380}$ && $-200_{-100}^{+100}$ &  \\
        Galaxies+QSO & $4_{-13}^{+13}$ && $770_{-400}^{+400}$ && $-186_{-100}^{+100}$ &  \\
        \hline
        Galaxies & $6_{-13}^{+13}$ && $-$ && $-$ & \\
        LRG only & $6_{-13}^{+14}$ && $-$ && $-$ & \\
         QSO-all ($P$-only)& $-2_{-10}^{+11}$ && $-$ && $-$ &\\
          QSO-all 
         & $-3_{-9}^{+9}$ && $-$ && $-$ &\\
           Galaxies+QSO-all+CMB & $-0.0_{-4.1}^{+4.1}$ && $-$ && $-$ & \\
           \hline
 Base (Galaxies+QSO), independent templates & $-$ && $690_{-400}^{+400}$ && $-165_{-100}^{+100}$ &  \\
Base ($P$-only)  & $-$ && $470_{-1600}^{+2200}$ && $1012_{-690}^{+520}$ &  \\
Base ($P$+$B_{\rm tree}$)  & $-$ && $610_{-430}^{+480}$ && $-63_{-140}^{+130}$ &  \\
Galaxies-hz+QSO  & $-$ && $746_{-440}^{+460}$ && $-170_{-120}^{+120}$ &  \\
Galaxies-hz+QSO+SBP  & $-$ && $-172_{-470}^{+400}$ && $-116_{-120}^{+120}$ &  \\
Galaxies-hz+QSO-all  & $-$ && $671_{-470}^{+410}$ && $-183_{-120}^{+120}$ &  \\
Galaxies-hz+QSO-all+SBP  & $-$ && $-98.4_{-410}^{+370}$ && $-123_{-120}^{+120}$ &  \\
        BOSS+SBP  & $-$ && $320_{-310}^{+290}$ && $100_{-140}^{+130}$ &  \\
        BOSS+DESI-hz+SBP  & 
        $-0.1_{-7.4}^{+7.4}$ && $200_{-230}^{+230}$ && $-24_{-86}^{+86}$ &  \\
        \hline
    \end{tabular}
    \caption{Constraints on local, equilateral, and orthogonal PNG templates from DESI DR1 power spectrum and bispectrum full-shape data. 
    \textit{First panel}: main results from the galaxy and quasar samples, including the high-redshift quasar sample `QSO-all'. 
    \textit{Second panel}: additional results for local PNG, including splits by tracer population, restricting to the QSO-all power spectrum alone (matching \citep{Chaussidon:2024qni}) and adding the \textit{Planck} PR4 constraints from \cite{Jung:2025nss}. 
    \textit{Third panel}: additional results for non-local PNG, including restriction to the tree-level bispectrum ($B_{\rm tree}$), addition of simulation-based priors (SBP), and limiting to high-redshift galaxies (galaxies-hz, or DESI-hz including QSO-all). Unless otherwise stated, we jointly vary $\fnl^{\rm equil}$ and $\fnl^{\rm orth}$, include both the power spectrum and bispectrum, and use conservative priors on EFT parameters.}
    \label{tab:params}
\end{table*}

\begin{figure}[!t]
\includegraphics[width=1\columnwidth]{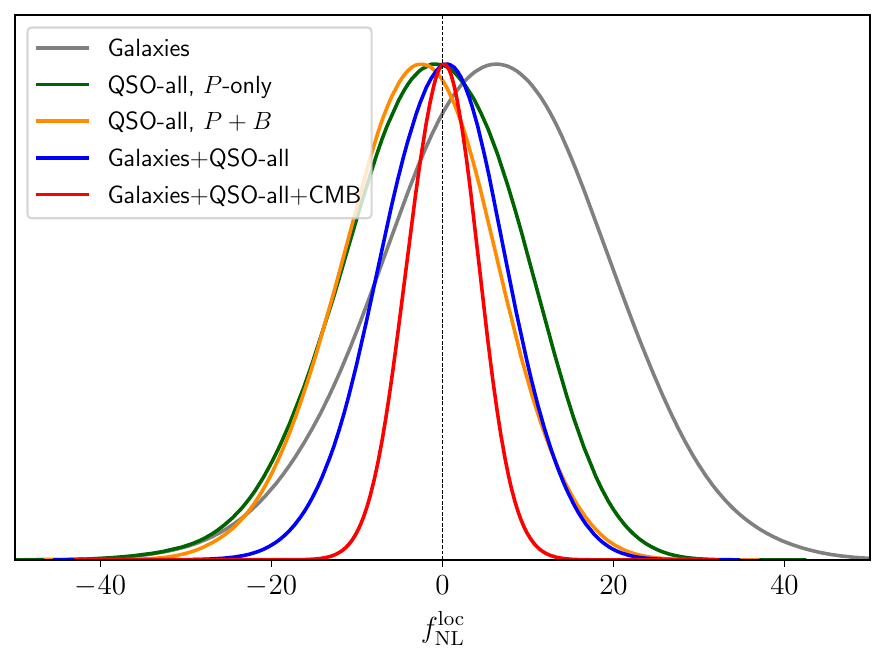}
\caption{Constraints on the local PNG amplitude, $f_{\rm NL}^{\rm loc}$, from various combinations of DESI galaxies and quasars. Much of our constraining power comes from the QSO-all power spectrum (which refers to the full DESI catalog up to $z_{\rm max}=3.1$), with constraints tightening when adding the bispectrum and other galaxy samples. The full DESI constraint (blue) is close to the CMB bound from \textit{Planck}; their combination (red) yields the tightest constraints on $f_{\rm NL}^{\rm loc}$ obtained to date. Numerical results are given in Tab.\,\ref{tab:params}.}
\label{fig:fnlloc}
\end{figure}

\begin{figure}
\includegraphics[width=1\columnwidth]{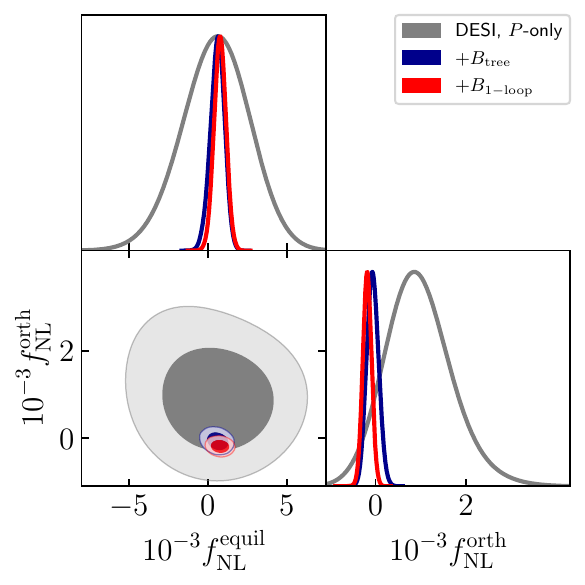}
\caption{Constraints on  single-field non-local PNG
amplitudes from the DESI galaxy and quasar clustering data. Whilst the constraints from the power spectrum (gray) are weak, we find improved results when adding the large-scale bispectrum (blue) and when extending to shorter scales with the one-loop theory (red). Numerical results are given in Tab.\,\ref{tab:params}.}
\label{fig:fnleq_desi}
\end{figure}

\begin{figure}[!t]
\includegraphics[width=1\columnwidth]{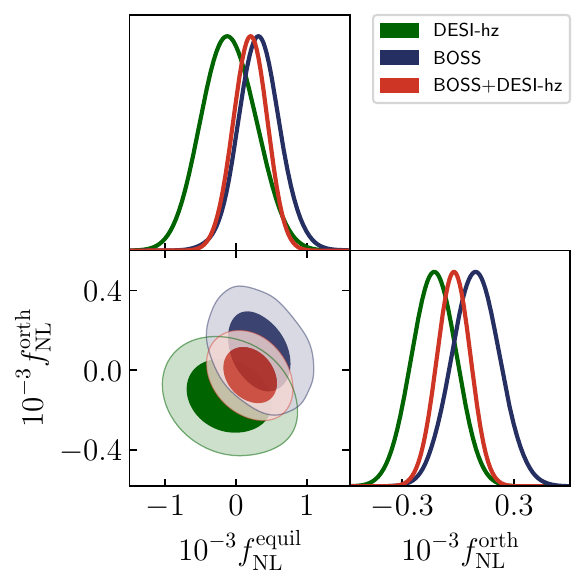}
\caption{Constraints on the single-field non-local PNG from DESI galaxy and quasar clustering data at high redshift, $z>0.6$ (DESI-hz), BOSS DR12, and their combination. Simulation-based priors are applied in all three analyses.}
\label{fig:fnleq_boss}
\end{figure}

\noindent Below, we present our constraints on both local and non-local PNG. Our main results are summarized in Tab.\,\ref{tab:params}.

\subsection{Local PNG}
\noindent In Fig.~\ref{fig:fnlloc} and the second panel of Tab.\,\ref{tab:params}, we display the local PNG measurements for various choices of dataset. Notably, the new tools and datasets used in our analysis lead to appreciably stronger constraints on local non-Gaussianity than those of the official DESI analysis~\cite{Chaussidon:2024qni}. In particular, we find $\fnl^{\rm loc}=6\pm 13$ from the DESI DR1 galaxies (without quasars), and $\fnl^{\rm loc}=6^{+14}_{-13}$ from the LRG samples alone. The latter bound is $\approx 10\%$ stronger than the analogous collaboration result $\fnl^{\rm loc}=2^{+15}_{-14}$ for the same choice $p=0.55$ and galaxy sample \citep{Chaussidon:2024qni}.
The addition of the QSO sample (with the default scale- and redshift-range) does not appreciably improve our bound. The same conclusion holds true 
for 
the BGS and ELG 
samples that were omitted 
in the official collaboration
analysis, for which
we find $\fnl^{\rm loc}=2_{-53}^{+57}$ from the BGS sample and $\fnl^{\rm loc}=-11_{-86}^{+71}$ for ELG2.
Although the addition of BGS1 and ELG2 samples does not
sharpen constraints appreciably, we do not find a significant reason to exclude them from the analysis.

Our galaxy constraints on local PNG are competitive with those from the optimized QSO-all sample. Using the QSO-all power spectrum only, we reproduce
the previous result $\fnl^{\rm loc}=-2^{+11}_{-10}$ \citep{Chaussidon:2024qni}; adding the large-scale QSO-all bispectum monopole improves this by about $15\%$, yielding $\fnl^{\rm loc}=-3\pm 9$. This is 
similar to the improvement found previously for eBOSS~\cite{Cagliari:2023mkq}. As an additional test, we verify that this constraint does not depend on the non-linear modeling. In particular, fitting the complete one-loop theory model to the power spectrum instead of the tree-level model, we obtain $\fnl^{\rm loc}=-6\pm 9$.

By combining the galaxy and QSO-all samples, we obtain the strongest LSS constraint on local PNG to date: $\fnl^{\rm loc}=0\pm 7$. Further combining this with the latest \textit{Planck}
CMB measurement (PR4; \citep{Jung:2025nss}) yields $\fnl^{\rm loc}=0\pm 4$. This is $18\%$ stronger than the CMB-only result, and represents the current most stringent bound on local PNG. 

\subsection{Non-Local PNG}
\noindent Next, we present constraints on equilateral and orthogonal PNG. Since the EFT of inflation generically predicts that both shapes are present in single-field scenarios \citep{Cheung:2007st}, we will mainly present results for a combined search for equilateral and orthogonal PNG, which are summarized in Figs.\,\ref{fig:fnleq_desi}\,\&\,\ref{fig:fnleq_boss} and the third panel of Tab.\,\ref{tab:params}. We also carry out single-template searches for each of the non-local shapes considered in this work, finding very similar results to those from the combined analyses due to a relatively weak correlations between the shapes. Results for the baseline single-template analysis are also shown in Tab.\,\ref{tab:params}.

We begin with the baseline DESI galaxy and quasar sample analyses with conservative EFT parameter priors, as shown in Fig.~\ref{fig:fnleq_desi}. We show results for three choices of likelihood: (1) power spectrum, (2) power spectrum and tree-level bispectrum (with $k_{\rm max}^{B_\ell}=0.08~\hMpc$), (3) power spectrum and one-loop bispectrum (which is the baseline choice). We observe that the bispectrum is \textit{crucial} for competitive constraints on the non-local PNG, since the the power-spectrum-only analysis yields $\fnl$ uncertainties about one order of magnitude larger than those coming 
from the tree-level bispectrum. Furthermore, we find that the one-loop bispectrum likelihood yields $\approx 12\%$ and $\approx 26\%$ improvements on $\fnl^{\rm equil}$ and $\fnl^{\rm orth}$ relative to the tree-level results. This is consistent with, though somewhat more modest than, the estimates for the one-loop bispectrum found in simulations~\cite{Philcox:2022frc,Bakx:2025pop}.
The relatively modest performance is attributed to the large number of free parameters in the one-loop EFT bispectrum, which cannot be well constrained with the DESI DR1 data, as well as the relatively small $k_{\rm max}$. 
Increase $k_{\rm max}$ leads to nominally stronger constraints~\cite{DAmico:2022gki}, though this comes at the expense of a large theory systematic error, which quickly surpasses the statistical one for $k\geq 0.17\,\hMpc$~\cite{Bakx:2025pop}.

In agreement with the modest gains from the one-loop bispectrum
quoted above, we find that the replacement of the base QSO sample (including the small-scale one-loop bispectrum) with the QSO-all sample (with only the large-scale tree-level bispectrum) improves over the baseline analysis by $3.5\%$,
giving $\fnl^{\rm equil}=719\pm 390$ and $\fnl^{\rm orth}=-200\pm 100$. Since this is nominally the strongest limit from DESI DR1, we adopt it as our primary constraint.

Our constraints on the equilateral and orthogonal shapes are $\approx 30\%$ stronger than those from the tree-level bispectrum analysis of SDSS-BOSS~\cite{Cabass:2022wjy,Ivanov:2024hgq,Reid:2015gra} ($\fnl^{\rm equil}=800\pm 500$ and $\fnl^{\rm orth}=8\pm 130$, see also \citep{DAmico:2022gki}), which were derived using conservative EFT parameter priors similar to those used above. In contrast, the simulation-based prior analysis of~\cite{Ivanov:2024hgq} found somewhat stronger BOSS constraints than those from our baseline DESI study, with $\fnl^{\rm equil}=320\pm 300$, $\fnl^{\rm orth}=100\pm 130$.

Given that the low-redshift $(z<0.6)$ DESI DR1 catalogs cover a smaller effective volume than BOSS DR12, one can combine BOSS and DESI to improve the constraining power. To do so, we replace the DESI BGS and LRG1 samples with the $z$1 and $z$3 BOSS DR12 samples analyzed in~\cite{Ivanov:2024hgq} (see~\cite{Chen:2024vuf} for a discussion of each sample).
The $z$3 sample has a partial overlap with DESI LRG2, but this is relatively small both in terms of the sky area ($\lesssim 70\%$~\cite{DESI:2024uvr}) and in the line-of-sight distribution (as seen by the different effective redshifts, with $z^{{\rm BOSS}-z3}_{\rm eff}=0.61$ vs $z^{\rm LRG2}_{\rm eff}=0.7$). As such, we ignore the overlap between these two samples as a first approximation. To compute a joint constraint, we combine the following samples: $z$1 and $z$3 from BOSS DR12, LRG2, LRG3, ELG2 and QSO-all from DESI DR1 (hereafter DESI-hz). 
We also perform analyses replacing QSO-all with QSO, which allows us to estimate the role of the small-scale one-loop bispectrum of quasars. 

When analyzing BOSS data, we use the tree-level bispectrum pipeline 
of~\cite{Ivanov:2024hgq} augmented with simulation-based priors (SBP)~\cite{Ivanov:2024hgq,Ivanov:2024xgb,Ivanov:2025qie,Chen:2025jnr}. For consistency, we also include SBP in the DESI-hz analysis. Due to the presence of the one-loop bispectrum, these improve the constraints quite marginally, with 
$\fnl^{\rm equil}=740\pm 450$, 
$\fnl^{\rm orth}=-170\pm 120$
in the DESI-hz conservative analysis (without BOSS, using the QSO quasar sample) and 
$\fnl^{\rm equil}=-170\pm 435$, 
$\fnl^{\rm orth}=-116\pm 120$
when SBP are included. 
When switching to the QSO-all sample, we find a slightly larger improvement ($\approx 12\%$ for $\fnl^{\rm equil}$): $\fnl^{\rm equil}=670\pm 440$, $\fnl^{\rm orth}=-183\pm 120$
(conservative) versus $\fnl^{\rm equil}=-100\pm 390$, $\fnl^{\rm orth}=-120\pm 120$ (SBP). We adopt the latter combination as the DESI-high-$z$ baseline for combining with BOSS. 

The combination of our DESI-hz likelihood and BOSS gives
$\fnl^{\rm equil}=200\pm 230$, $\fnl^{\rm orth}=-24\pm 86$,
which yields $\approx 20\%$ and $\approx 34\%$ improvements
on the respective templates with respect to the BOSS-only results. While our limit on the equilateral 
non-Gaussianity
is the strongest one derived from LSS, it is still somewhat weaker than WMAP-Y9 result $\fnl^{\rm equil}=51\pm 136$~\cite{WMAP:2012fli}.
However, our new constraint on the orthogonal shape is already $\approx 14\%$ stronger than that of WMAP-Y9, $\fnl^{\rm orth}=-245\pm 100$. Both constraints,
however, 
are significantly weaker than those from \textit{Planck} \citep{Jung:2025nss,Planck:2019kim}.

\section{Conclusions}
\label{sec:concl}

\noindent
We have presented a search for primordial non-Gaussianity in the first-year DESI data, using the power spectra and bispectra of galaxies and quasars. Our analysis features several important improvements with respect to previous works. Most notably, we include the DESI bispectrum, which is made possible by our efficient estimator, careful treatment of systematic effects, and the development of a consistent theoretical model for the one-loop galaxy bispectrum, accounting for non-linear galaxy bias, redshift-space distortions, and galaxy stochasticity on quasi-linear scales.

The use of the full DESI data, including the high-redshift quasar power spectrum and bispectra, has resulted in the strongest constraints to date on local PNG
in our work, surpassing both the official collaboration limits \citep{Chaussidon:2024qni} and (in combination) those from \textit{Planck}. In addition, we have performed a search for non-local PNG
in DESI data for the first time, finding $\lesssim 30\%$ improvements over analogous studies using BOSS data. Significant improvements in bispectrum estimation and modeling in our current analysis pipeline substantially improve the robustness of the reported constraints on non-local PNG shapes in comparison with previous BOSS results, as discussed in \cite{Chudaykin:2024wlw,Bakx:2025pop}.

Our analysis can be extended in a number of ways. From a phenomenological perspective, it will be interesting to analyze other well motivated theoretical 
templates for PNG, including cosmological collider
models~\cite{Chen:2009zp,Chen:2009we,Chen:2010xka,Baumann:2011nk,Arkani-Hamed:2015bza,MoradinezhadDizgah:2017szk,MoradinezhadDizgah:2018ssw,MoradinezhadDizgah:2019xun,MoradinezhadDizgah:2020whw,Kumar:2018jxz,Kumar:2019ebj,Reece:2022soh,Cabass:2022oap,Cabass:2024wob}, as well dissipative and warm inflation models~\cite{Mirbabayi:2022cbt,Salcedo:2024smn}.
It will also be interesting to study
the sensitivity
of our analysis
to the simulation-based
priors more systematically in
order to identify physical 
effects in the halo occupation distribution
that are primarily
correlated 
with PNG.

Our work establishes the galaxy bispectrum as a robust observable to extract early Universe signals from the observed distribution of galaxies. This provides a unique tool to test physics that might have operated at the highest energies accessible to experiment; harnessing it will be essential to enable transformative discoveries in early Universe cosmology.

\vskip 8pt
\acknowledgments
{\small
\begingroup
\hypersetup{hidelinks}
\noindent 
AC acknowledges funding from the Swiss National Science Foundation. The computations in this work were run at facilities supported by the Scientific Computing Core at the Flatiron Institute, a division of the Simons Foundation, as well as at the Helios cluster at the Institute for Advanced Study, Princeton. OHEP thanks the \href{https://www.flickr.com/photos/198816819@N07/54960424322}{Parisien animallia} for support whilst this work was being completed.

\noindent This research used data obtained with the Dark Energy Spectroscopic Instrument (DESI). DESI construction and operations is managed by the Lawrence Berkeley National Laboratory. This material is based upon work supported by the U.S. Department of Energy, Office of Science, Office of High-Energy Physics, under Contract No. DE–AC02–05CH11231, and by the National Energy Research Scientific Computing Center, a DOE Office of Science User Facility under the same contract. Additional support for DESI was provided by the U.S. National Science Foundation (NSF), Division of Astronomical Sciences under Contract No. AST-0950945 to the NSF’s National Optical-Infrared Astronomy Research Laboratory; the Science and Technology Facilities Council of the United Kingdom; the Gordon and Betty Moore Foundation; the Heising-Simons Foundation; the French Alternative Energies and Atomic Energy Commission (CEA); the National Council of Humanities, Science and Technology of Mexico (CONAHCYT); the Ministry of Science and Innovation of Spain (MICINN), and by the DESI Member Institutions: \url{www.desi.lbl.gov/collaborating-institutions}. The DESI collaboration is honored to be permitted to conduct scientific research on I’oligam Du’ag (Kitt Peak), a mountain with particular significance to the Tohono O’odham Nation. Any opinions, findings, and conclusions or recommendations expressed in this material are those of the author(s) and do not necessarily reflect the views of the U.S. National Science Foundation, the U.S. Department of Energy, or any of the listed funding agencies.
\endgroup
}

\bibliography{short.bib}

\end{document}